\documentclass[preprint2]{aastex}

\usepackage{amsmath, amssymb, amsfonts}

\shortauthors{Sartore et al.}

\begin{document}


\title{The {\it INTEGRAL}/SPI view of A0535+26 during the giant outburst of February 2011.}


\author{N. Sartore\altaffilmark{1}, E. Jourdain\altaffilmark{1,2}, and J.~P. Roques\altaffilmark{1,2}}
\affil{CNRS - IRAP, 9 Av. Colonel Roche. BP 44346,F-31028 Toulouse cedex 4, France}
\affil{Universit\'e de Toulouse, UPS-OMP, IRAP, Toulouse, France}

\email{nicola.sartore@irap.omp.eu}

\begin{abstract}
A0535+26 is a slowly rotating pulsar accreting from the wind of a massive Be star, and that exhibits
two cyclotron absorption lines in its X-ray spectrum, at about 45 and 100 keV, respectively. 
Unlike similar sources, no significant variations of the energy of its cyclotron lines with flux were observed to date.
The bright outburst of February 2011 thus offers a unique occasion to probe this peculiar behavior at 
flux levels not yet observed with present-day instruments.
Here we report on the spectral and timing analysis of the data from the spectrometer SPI on-board {\it INTEGRAL} 
collected during the outburst. 
At the peak of the outburst the estimated luminosity is $\sim4.9\times10^{37}\,\rm erg\,s^{-1}$.
The fundamental cyclotron feature is detected at all flux levels, and its centroid energy is positively-correlated with 
the flux of the source, confirming that A0535+26 is accreting at a sub-critical regime. 
The correlation seems to fall off at $\sim10^{37}\,\rm erg\,s^{-1}$, suggesting the transition from a Coulomb-stopping 
regime to a gas-mediated shock regime.
From the timing analysis we found that the pulsar was spinning up during most of the outburst, and that the spin-up rate correlates
with the flux of the source, albeit the correlation is steeper than the one expected from the standard disk accretion theory.
Finally, we show that the pulse profile of the source changes dramatically as the flux increases. 
At high luminosity the profile is highly asymmetric, implying an asymmetry in the geometry of the accretion flow.
\end{abstract}


\section{INTRODUCTION}\label{sect-intro}
High-mass X-ray binaries (HMXB) are systems composed by a massive, early type star which provides the matter to
an accreting compact object, typically a highly-magnetized neutron star, $B\sim\,10^{11} - 10^{13}\, \rm G$.
This has been inferred from the presence in their X-ray spectra of absorption-like features, which are interpreted as 
cyclotron resonant scattering features (CRSF), where $E_{CRSF}\sim11.6\,(B / 10^{12}\, \rm G)$ keV, and {\it B} is the 
magnetic field of neutron star.
Such strong magnetic fields channel the accretion flow towards the magnetic poles of the neutron star.
The resulting anisotropy induces a modulation in the radiation emitted by the neutron star, which is then observed as an 
X-ray pulsar \citep[see e.g.][for recent reviews]{wilms2014,revnivtsev2014}.

The accreted matter may come either from an accretion disk, via Roche lobe overflow, or directly from the wind of the 
massive companion. 
Wind-fed HMXBs are further sub-divided into two groups, super-giant and Be systems, depending on the properties of 
the donor star. 
In particular, Be stars are fast-rotating OB stars in which the outflow takes the form of a dense equatorial disk, 
as suggested by observations of broad emission lines at optical/infrared wavelengths \cite[e.g.][]{porter2003}.
Be/HMXB exhibit long periods of quiescence, interrupted by bursts of activity which are classified in two groups 
\citep{stella1986}.
The so-called type I, or normal, outbursts occur quasi-periodically at orbital phases close to the periastron passage, 
and reach a luminosity of $\sim10^{37}\, \rm erg\,s^{-1}$ at most. They are likely related to an increase in the 
accretion rate due to the passage of the pulsar through the equatorial disk of the companion.
On the other hand, type II (giant) outbursts are less frequent, with no apparent correlation with the orbital phase, 
and can reach a luminosity close to the Eddington limit\footnote{At this luminosity, the force exerted by the 
emerging radiation on the in-falling matter through Compton scattering balances the gravitational force.}, 
$L_{\rm Edd} \sim 1.38\times10^{38}\,(M/M_{\odot})\,\rm erg\,s^{-1}$, where $M$ is the mass of the neutron star. 
The origin of type II outbursts is not clear and still matter for debate \citep[see][and references therein]{nakajima2014}.

Interestingly, these changes in luminosity can be used to test the accretion theory on strongly-magnetized pulsars at 
different accretion rates.
In accreting pulsars a radiation-dominated shock forms in the accretion column at a luminosity $L_X \gtrsim 10^{36}\,\rm erg\,s^{-1}$ 
\citep{basko1976}.
As the matter passes through the shock, it gives away most of its kinetic energy, which is transferred to the radiation field.
Below the shock, the accreted matter slowly sinks towards the surface of the pulsar where it can be further decelerated. 
The final deceleration depends on the luminosity. For a certain critical value $L_{crit}$, which depends on the properties of the 
neutron star, like e.g. the mass and the strength of magnetic field \citep[see Equation 32 of][]{becker2012}, the pressure exerted 
by the upward radiation flux is such that the accretion flow arrives at the surface of the pulsar with zero speed.
Above this critical value the dynamics of the accreted matter is dominated by radiation pressure, and both the radiative shock 
and the line forming region are expected to move upwards where the magnetic field is lower. 
This in turn implies an anti-correlation between the energy of the cyclotron line and the luminosity of the source, 
as reported e.g. for the Be/HMXB systems 4U 0115+63 and V0332+53 \citep[see e.g.][]{nakajima2006, tsygankov2006, mowlavi2006}.

On the other hand, in sources below the critical luminosity, the radiation pressure cannot stop completely the in-falling 
matter, which is instead decelerated by Coulomb interactions towards the base of the accretion column \citep{burnard1991}. 
In this case, the height of the region where the deceleration occurs decreases with luminosity 
\citep[see][for details]{becker2012} and, because of the increase of magnetic field at lower altitudes, a positive correlation 
between the energy of the cyclotron line and the flux is then expected.
Such behavior has been observed e.g. in Her X-1, GX 304$-$1, {\it Swift} J1626.6$-$5156, and Vela X-1 
\citep{staubert2007, klochkov2012, decesar2013, furst2014}.

\subsection{The case of A0535+26}
A0535+26 (A0535 hereafter) is a transient Be/HMXB discovered by the {\it Ariel V} satellite during a giant outburst 
\citep{rosenberg1975}.
The systems lies at a distance $d\sim 2\,\rm kpc$ \citep{steele1998} from the Earth, and it is composed by a slowly 
rotating pulsar, $P_{spin}\sim103$ seconds, orbiting the B0IIIe star HDE 245770 in an eccentric orbit ($e=0.47\pm0.02$), 
with a period $P_{orb} = 111\pm0.1$ days \citep{finger2006}.
After its discovery, A0535 exhibited several bursting episodes of both types I and II. The hard X-ray spectrum of the source 
during these outbursts showed two prominent absorption lines at about 45 and 100 keV 
\citep[e.g.][and references therein]{caballero2013}.

At variance with the examples reported above, the energy of the fundamental cyclotron line in A0535 was found to be 
remarkably stable at $\sim45\,\rm keV$ in almost all observations of the source. The only exception occurred during a 
short flaring episode in which a higher value, $\sim52\,\rm keV$, was reported \citep{caballero2008}. 
This increase in the observed cyclotron energy was coupled with changes of the pulse profile, and was explained by the 
occurrence of instabilities in the magnetosphere of the pulsar leading to a sudden increase in the accretion rate and 
changes in the geometry of the accretion column \citep{postnov2008}.
According to \citet{klochkov2011}, and \citet{muller2013}, A0535 exhibits a positive correlation of the cyclotron energy 
with the luminosity on a pulse phase-dependent basis. While this result suggests that the source is in the sub-critical 
regime, a definitive proof of the correlation of the cyclotron lines with the luminosity in A0535 is still lacking.

Dedicated observations of A0535 during a giant outburst are thus important in order to extend the range of luminosity 
covered by spectroscopic studies of the source.
In this regard, the two previous giant outbursts (May/June 2005 and December 2009) were not well suited for the task due 
to Sun constraints, i.e. the source was too close to the Sun in the sky to be safely observed.
On the other hand, the giant outburst of February 2011 was well covered by several instruments, and it offers a unique 
possibility to reveal changes, or lack thereof, in the spectral properties of A0535 and their relation with the observed 
luminosity, thus probing the accretion regime of the source.
An analysis of {\it Swift}/BAT and {\it INTEGRAL}/ISGRI data below 80 keV taken during the outburst has been presented by 
\cite{caballero2012}, where they reported that the energy of the fundamental cyclotron line was stable on all but one 
of the {\it INTEGRAL}/ISGRI observations.

In this paper, we report on the observations of the Be/HMBX A0535 during the giant outburst of February 2011, 
performed with the SPI instrument on-board {\it INTEGRAL}. 
In Section \ref{sect-obs-data}, we give a description of SPI, as well as of the observations and the data reduction procedure 
used in the paper. 
In Section \ref{sect-results}, a pulse phase-averaged spectroscopic analysis is presented, completed by a timing analysis and 
a qualitative study of the energy-resolved pulse profiles.
Finally, in Section \ref{sect-discuss}, we discuss our results and present the conclusions.

\section{OBSERVATIONS AND DATA REDUCTION}\label{sect-obs-data}
{\it INTEGRAL} \citep{winkler2003} is a mission of the European Space Agency devoted to the observation of the high-energy 
sky (from $\sim\rm keV$ to $\sim\rm MeV$). The satellite was launched from Baikonur in October 2002 and put in a highly 
eccentric orbit. The spectrometer SPI (SPectrometer on Integral), operating between 20 keV and 8 MeV, is one of the two main 
instruments on-board {\it INTEGRAL}. 
A detailed description of the instrument and its in-flight performances can be found in \citet{vedrenne2003} and 
\citet{roques2003}. It consists of 19 high purity Germanium detectors maintained at a temperature of $\sim$80 K, 
which ensures an excellent energy resolution all along the allowed energy range. 
The instrument uses a tungsten-made coded-mask offering a 2.6$^\circ$ angular resolution over a 30$^\circ$ field of view 
(f.o.v.). 
Particle and photon background reduction is achieved thanks to a 5 cm Germanate Bismuth anti-coincidence shield.

{\it INTEGRAL} observations are carried-out on a revolution-by-revolution basis, each revolution (rev. in short) 
lasting $\sim$3 days, of which $\sim2.5$ days give useful data. 
Each observation is further sub$-$divided into shorter, $\sim$2-4 kiloseconds exposures called ``science windows'' (scw).
The low number of pixels enforces the adoption of a dithering procedure \citep{jensen2003}, in order to better estimate 
the contributions of the background and of the different sources in the f.o.v. 
The satellite aiming point is thus shifted by $\sim2^\circ$ around the observed target at each scw, 
following a predefined pattern, typically a $5^\circ\times5^\circ$ rectangle.

\begin{figure}[b]
\epsscale{1.0}
\plotone{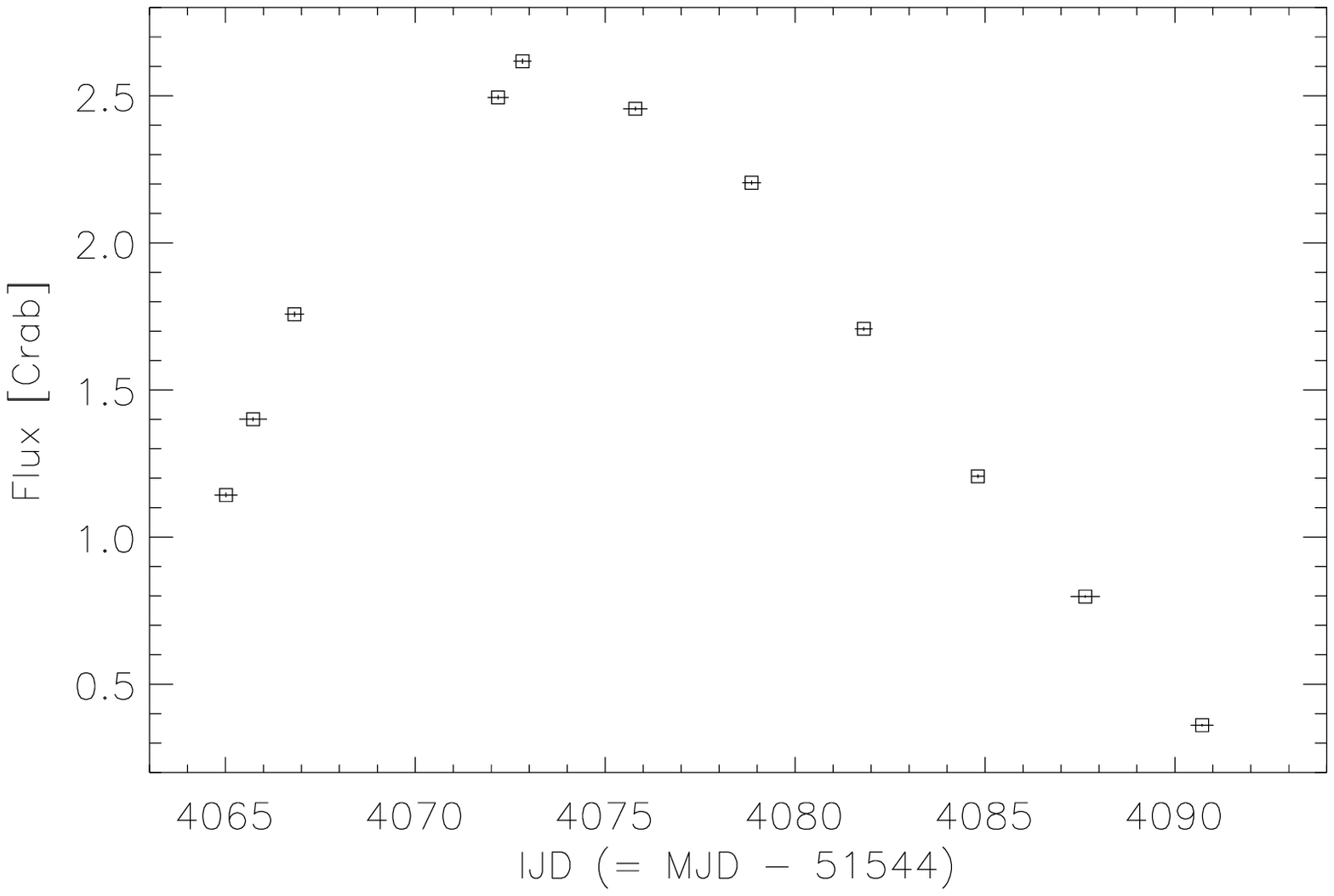}
\caption{Evolution of the of 20-80 keV flux of A0535 during the outburst of February 2011.}\label{fig_flux_vs_time}
\end{figure}

The giant outburst of February 2011 of A0535 has been monitored in {\tt ToO} (Target of Opportunity) mode by {\it INTEGRAL} 
in 8 consecutive revolutions (rev. 1021 to 1028), which caught essentially the maximum and the decline of the outburst.
The total exposure of each observation varied between $\sim$40 and $\sim$100 kiloseconds. 
In addition, the rise of the outburst was observed serendipitously during rev. 1019, a $\sim186$ kilosecond-long 
calibration observation of the nearby Crab Nebula (Figure \ref{fig_flux_vs_time}). 
A log of the {\it INTEGRAL} observations of A0535 used in this paper is reported in Table \ref{table_log}.

The spectrum of each source in the f.o.v. of SPI is evaluated by means of a sky model fitting procedure 
\citep[see][for a detailed description]{jourdain2009}. 
In the selected observations, the f.o.v. always contains three sources: A0535, the Crab Nebula, and the low-mass X-ray 
binary 4U 0614+091. However, the latter is much fainter than the former two, so we can safely neglect it from 
our sky model.
For each revolution, we extract the background-subtracted spectra of A0535 and of the Crab, together with the corresponding 
response matrices with SPIDAI\footnote{SPI Data Analysis Interface. 
The software has been developed at the IRAP Toulouse, and it is available on the web at {\tt http://sigma-2.cesr.fr/integral/spidai}. 
For more details, see \cite{burke2014}.}. 
We selected spectral counts in the 20-135 keV range, and divided them into 45 logarithmically-spaced bins.
For the evaluation of the spectral counts from each source we assume a variation time-scale of one scw for the flux of A0535, 
while for the Crab we assume a constant flux over the time-scale of the observation.
We exclude scw affected by high particle background rates, e.g. in case of solar flares or close to the satellite 
ingress/egress from the Earth radiation belts.

Thanks to their long exposures, we are able to split rev. 1019 in three sub-sets (labeled A, B, and C, respectively),  
and rev. 1021 in two sub-sets (labeled respectively A and B), in order to check for variations of the spectral and 
timing parameters within the revolution time-scale. 

We also extract event lists corresponding to each data sub-set in order to follow the evolution of the pulsar spin period and 
of the associated pulse profiles.

\section{RESULTS}\label{sect-results}

\subsection{Spectral Analysis}\label{sect-results-spectra-rep}
The analysis of the source spectra has been performed with XSPEC 12.8.2 \citep{arnaud1996}. 
The hard X-ray emission of accreting pulsars is thought to originate from the thermal and bulk comptonization of seed photons
in the accretion column \citep[see e.g.][for details]{becker2007}.
The resulting spectral shape can be well approximated by a power law continuum with exponential cut-off at high energies 
({\tt cutoffpl} in XSPEC), $F(E)\,=\,A\,E^{-\Gamma}\,\exp(-E/E_{cut})$, where $\Gamma$, $E_{cut}$, and $A$ are respectively 
the spectral index of the power law, the folding energy and the normalization constant of the model spectrum, 
and $E$ is the energy of the incoming photons in keV. 
We first perform a fit of each spectrum with this model. 
We notice that the spectrum of rev. 1028 has low count statistics, hence we exclude it from the spectral analysis.

\begin{figure}[b]
\epsscale{1.0}
\plotone{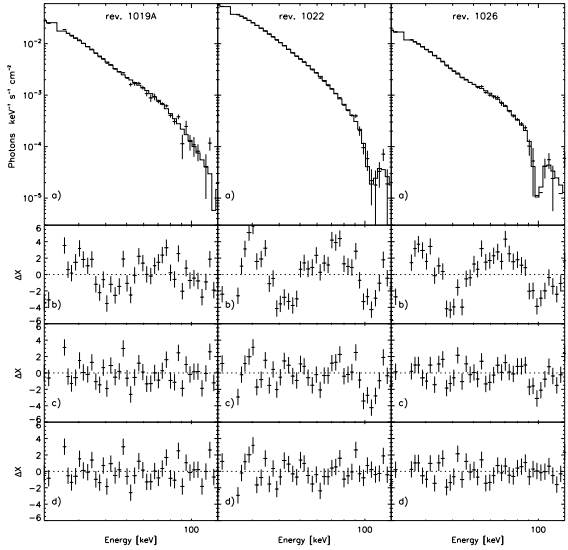}
\caption{Spectra (top panel) and associated fit residuals (second to fourth panel from top to bottom) of A0535 at different epochs of the outburst, 
and corresponding, from left to right, to rev. 1019A, rev.1022, and rev. 1026, respectively. }\label{fig_spectra}
\end{figure}

All spectra show large structured residuals around $\sim45\,\rm keV$ (see Figure \ref{fig_spectra}), pointing out the presence 
of the fundamental cyclotron feature in our data.
We account for this feature by including a multiplicative absorption line with Gaussian profile ({\tt gabs} in XSPEC) 
in the spectral model, which becomes $F'(E)=F(E)\exp[-\tau(E)]$, 
where $\tau(E)=(\tau_{cyc}/\sqrt{2\pi}\sigma_{cyc})\exp[-(E-E_{cyc})^2/2\sigma_{cyc}^2]$ and 
$E_{cyc}$, $\sigma_{cyc}$, and $\tau_{cyc}$ are respectively the centroid energy, the width, and the depth of the 
cyclotron line. This results in significant improvements in the $\chi^2$ in all cases.

\begin{figure}[b]
\epsscale{1.0}
\plotone{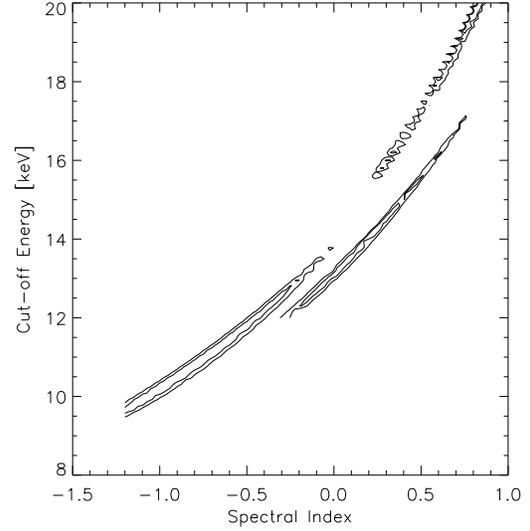}
\caption{Contour plot of the cut-off energy $E_{cut}$ versus the spectral index $\Gamma$.}\label{fig_continuum_degen}
\end{figure}

We notice that there is a strong coupling between the spectral index and the cut-off energy of the continuum model 
(Figure \ref{fig_continuum_degen}).
In addition, the width of the absorption line is anti-correlated with parameters of the continuum. 
This may result in unrealistic values of the fit parameters.
For this reason, we fix the spectral index to an average value, $\Gamma = 0.5$ \citep[e.g.][]{muller2013} and repeat the fits,

To assess the significance of the absorption lines we proceeded as follows.
First, for each data-set we generate 10000 simulated spectra with the in-built {\tt fakeit} command from XSPEC, using the 
best-fit model as a template.
Then we fit each simulated spectrum with either the {\tt cutoffpl} or the {\tt cutoffpl * gabs} models, 
imposing that the energy of the line is equal to the corresponding best-fit value, and taking note of the resulting $\chi^2$.
From these we build the distribution of the {\it F} statistics, defined as $F=(\chi_0^2/\nu_0)/(\chi_1^2/\nu_1)$ 
\citep[see][and references therein]{orlandini2012}, where $\nu_i$ is the number of degrees of freedom of model {\it i} and 
the subscripts $i=0,1$ correspond respectively, to the models without and with the absorption line.
Finally we infer the probability of a chance improvement of the $\chi^2$ by counting how many times the simulated 
values of {\it F} were larger than that obtained from the real data. 
In all cases the estimated chance probability is lower than $10^{-4}$, implying a significance greater than $3.9\sigma$.

From part B of rev. 1021 onward the spectrum exhibits structured residuals at $\sim100\,\rm keV$.
The existence of a second cyclotron line at $\sim100$ keV in the spectrum of this source has been reported several times 
in the past \citep[see e.g.][]{caballero2007}, and the residuals observed in our spectra are likely due to this feature.
As in the case of the line at lower energy, we perform simulations in order to assess the significance of
these absorption lines, where we now include the fundamental feature at $\sim 45\, \rm keV$ in the null hypothesis.
The line is formally significant ($>3\sigma$) for rev. 1021B, 1022, 1023, and 1026, while for rev. 1024 the significance is barely 
$\sim2.3\sigma$. We nevertheless include the second line in the spectral model for all these observations.
Finally, for rev. 1025 and 1027 we find no evidence of residuals around 100 keV, possibly due to a lack of statistic caused by the 
combined effect of the fading flux of the source and of the short exposure times.
We report the results of the spectral fitting in Table \ref{table_specfit}, where all the uncertainties are at the 
$1\sigma$ confidence level and have been estimated by means of simulations.

For each spectrum, we estimate the corresponding flux in the $20-80$ keV range, and normalize it to that of the Crab Nebula 
obtained simultaneously from the same observations (Figure \ref{fig_flux_vs_time}).
The flux reaches a maximum of $\sim3.87\times10^{-8}\,\rm erg\,cm^{-2}\,s^{-1}$ ($\sim2.6$ Crab) during rev. 1021B. 
This corresponds, at a distance of $\sim2$ kpc, to a luminosity $L_{bol}\sim4.9\times10^{37}\,\rm erg\,s^{-1}$, 
where we assumed isotropic emission and a bolometric correction factor of 40\% \citep{bildsten1997}. 

\begin{figure}[b]
\epsscale{0.8}
\plotone{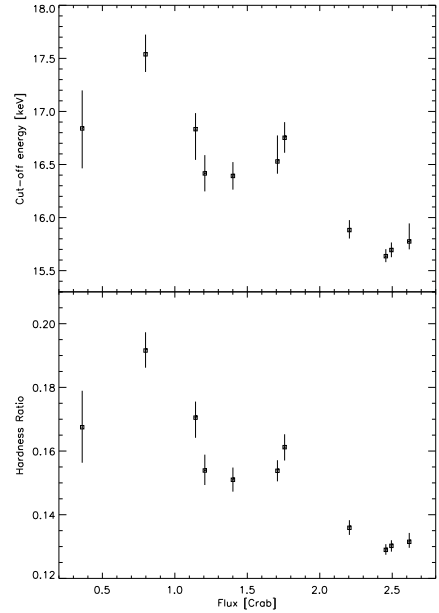}
\caption{({\it Top}) Relation between the best-fit cut-off energy of the continuum and the 20-80 keV flux.
({\it Botton}) Relation between the hardness ratio of the spectrum and the flux of the source. The 
soft and hard fluxes have been estimated in the 20-25, and 70-80 keV bands, respectively. 
Error bars represent the $1\sigma$ uncertainties.
}\label{fig_spec_vs_flux}
\end{figure}

We then look for possible correlations of the spectral parameters with the observed flux, in order to constrain the accretion 
regime of the source.
We find that the cut-off energy of the spectral continuum is anti-correlated with the flux (Figure \ref{fig_spec_vs_flux}, upper panel). 
Accordingly, the spectrum softens as the flux increases, as showed by the hardness ratio between fluxes in the $20-25$ and $70-80$ 
keV bands (lower panel of Figure \ref{fig_spec_vs_flux}). 
The boundaries of the two energy bands were chosen in order to limit as much as possible the contribution of the two cyclotron 
lines.

\begin{figure}[b]
\epsscale{1.0}
\plottwo{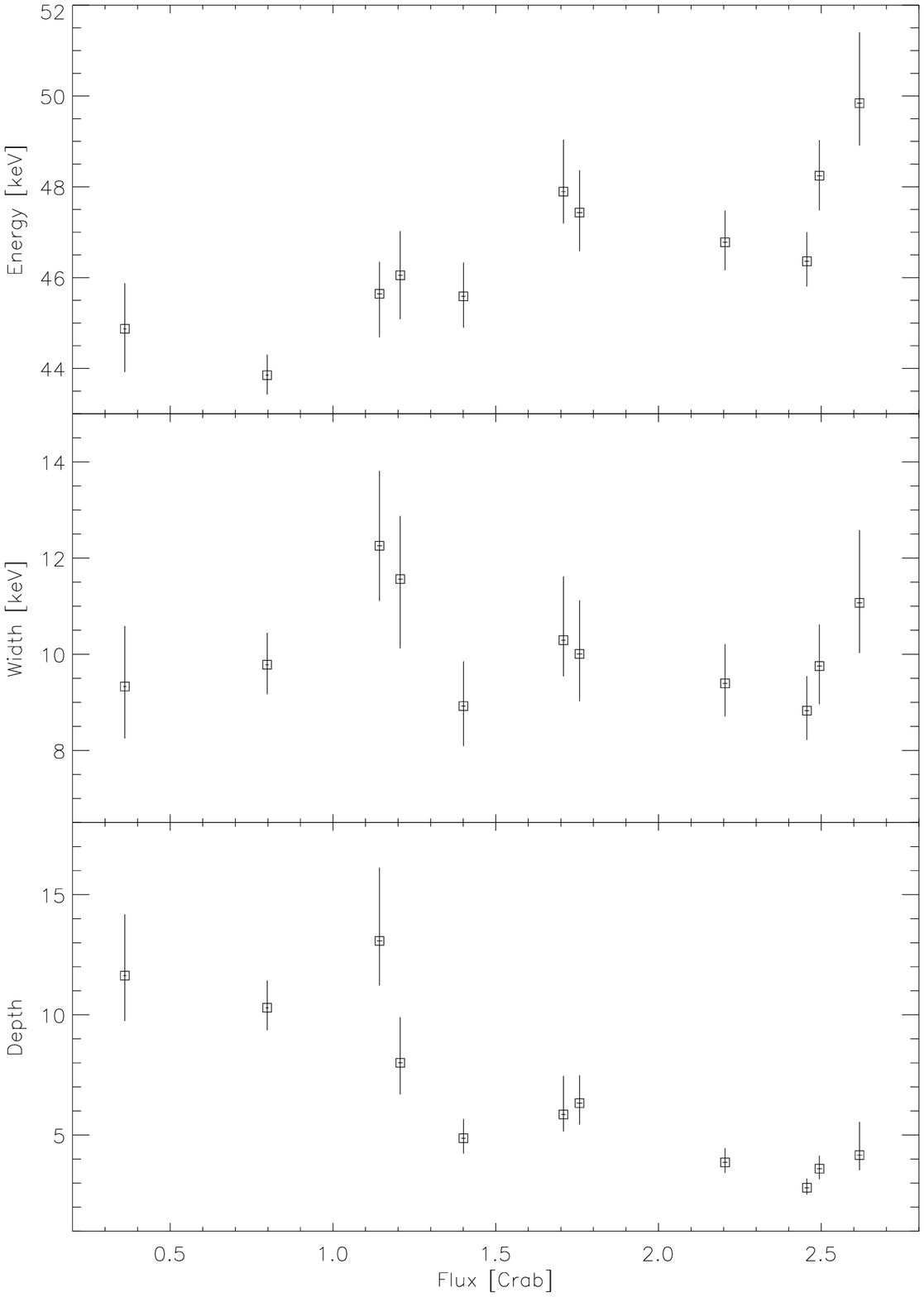}{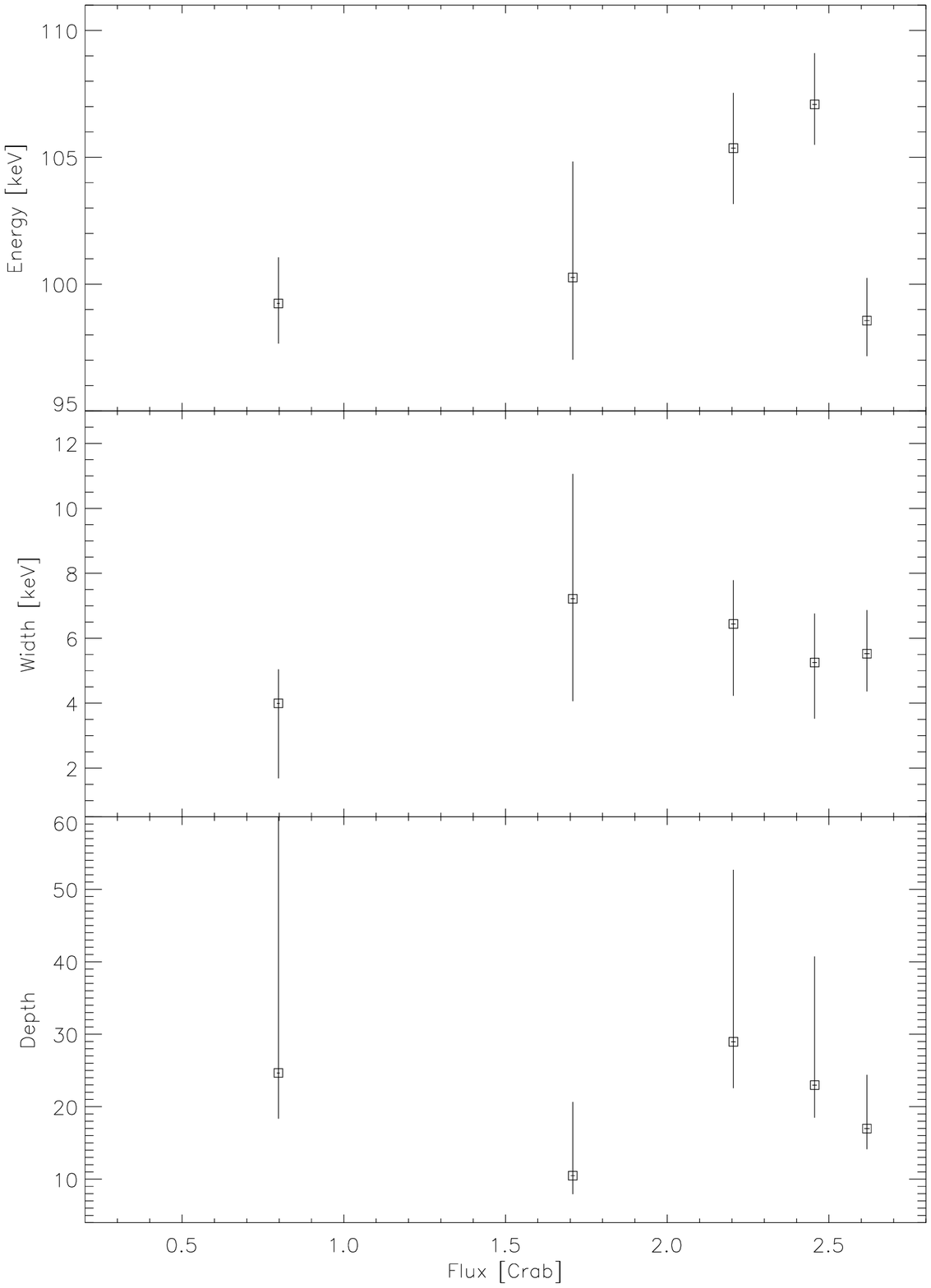}
\caption{Relation between the best-fit parameters and the 20-80 keV flux of A0535 for the cyclotron lines 
at $\sim45$ keV (left), and $\sim100$ keV (right). The top, middle and bottom panels show respectively the centroid energy, 
the width and the depth of the lines. Error bars represent the $1\sigma$ uncertainties.
}\label{fig_lines_vs_flux}
\end{figure}

Intriguingly, the relations between the parameters of the fundamental cyclotron line and the flux of the source differ between each 
other (Figure \ref{fig_lines_vs_flux}, left panels). The centroid energy $E_{cyc_1}$ is positively-correlated with the flux of A0535, 
going from $\sim50$ keV at high luminosity (rev. 1021B) to $\sim44$ keV at low luminosity (rev. 1026). 
We notice that the energy of the line at the lowest luminosity (rev. 1027) is compatible with that from the previous revolution, 
which implies a possible change in the correlation at about this luminosity.

\begin{figure}[b]
\epsscale{1.0}
\plotone{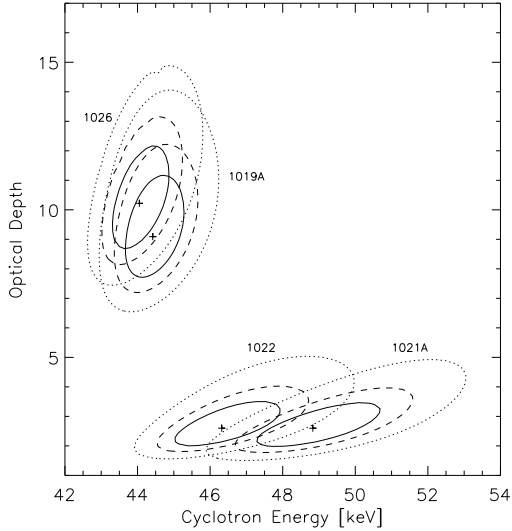}
\caption{Contour plot of the line depth $\tau_{cyc_1}$ versus the centroid energy $E_{cyc_1}$ of the fundamental cyclotron line.
Countour levels are drawn at $\chi_{min}^2 + 2.30$ (solid), $\chi_{min}^2 + 4.61$ (dashed), $\chi_{min}^2 + 9.21$ (dotted), respectively,  
and represent the $1\sigma$, $2\sigma$, and $3\sigma$ confidence intervals for the two parameters of interest.}\label{fig_line1_contours}
\end{figure}

On the other hand, the width of the line $\sigma_{cyc_1}$ remains in the $9-13\, \rm keV$ range, and it does not exhibit a correlation with 
the flux,  while the line depth $\tau_{cyc_1}$ presents a clear anti-correlation with the flux (see Figure \ref{fig_lines_vs_flux}), 
going from $\sim13$ at low luminosity to $\sim3$ at high luminosity. 
Figure \ref{fig_line1_contours} shows the contour plot for the optical depth $\tau_{cyc_1}$ versus the centroid energy $E_{cyc_1}$
of the fundamental cyclotron line at the beginning (rev. 1019A), maximum (rev. 1021A and 1022), and towards the end of 
the outburst (rev. 1026). 
It can be observed that the depth of the line exhibits significant variations during the outburst, 
while the variations of the centroid energy are less significant.
The parameters of the line at the end of the outburst are very close to the ones at the beginning with similar flux level, 
suggesting the absence of hysteresis effects during the outburst.

Finally, we find no correlation between the flux of the source and the parameters of the first harmonic.
However, the energy of the line seems to increase from $\sim100$ keV to $\sim108$ keV between rev. 1021B and 1022, 
and to decrease again to $\sim100$ keV from rev. 1024 onward even though the lack of statistics prevents any firm conclusion.

\subsection{Timing analysis}\label{sect-results-timing}
For each event list, we first convert photon arrival times to the Solar System barycenter and apply corrections 
to account for the binary motion of A0535. Ephemerids for the binary orbit are taken from \cite{finger1996}, 
adopting the updated values from \cite{camero2012} for the epoch of the periastron passage and the orbital period. 

We then carry out a $Z_2^2$ test \citep{buccheri1983} around the known pulse frequency of the source on each event list.
For this task, we select only photons in the $24-49$ keV range, where the signal-to-noise ratio is higher.
To further increase the signal-to-noise ratio, we use the information from the spectral deconvolution algorithm: 
for all the available scw, we estimate the fraction $\alpha$ of source photons that reach each detector due to the shadow casted by 
the coded mask on the detectors, and select only photons from detectors with $\alpha$ larger than a given value 
\cite[see][for details]{molkov2010}. In this work, we exclude events from detectors with $\alpha < 0.2$.

The $Z_2^2$ periodogram exhibits a very large peak around the reported spin period of the pulsar in all data-sets, 
with values of the power ranging from $\sim6\times10^2$ to $\sim1.5\times10^4$ against an expected average of 4.
Following \cite{leahy1987}, we fit the periodograms around the peak of the $Z_2^2$ distribution with a Gaussian model in order 
to better estimate the spin frequency. 
The $1\sigma$ uncertainties on the best-fit frequencies were inferred by adapting equation B17 of \cite{ransom2002} 
to the $Z_2^2$ case.

Then, we perform phase-connected timing in order to determine the spin evolution of the pulsar during the outburst.
The pulse phase at a time {\it t} can be expressed as:

\begin{equation}
 \phi(t) = \phi_0 + f_0(t-t_0) + \frac{1}{2}\dot{f_0}(t-t_0)^2 + ...
\end{equation}

\noindent where {\it f} is the spin frequency, the dots represent derivation with respect to the time, and the subscript 0 denotes 
quantities evaluated at the reference epoch. We take ${\rm MJD} = 55616.202$, the mid-point of rev. 1021A, as the reference epoch 
$t_0$.
Starting from a simple constant frequency model, $f(t) \equiv f_0$, for each data-set, we produce pulse profiles using events in the 
$27 - 49\, \rm keV$ range, and measure the phase shift as a function of time with respect to the reference epoch.
The phase is measured at the minimum of the pulse profile.
We find a reasonable fit using a polynomial model which includes the derivatives of the frequency up to the fourth order. 
The best-fit coefficents of the timing solution are reported in Table \ref{table_timing_solution}.

\begin{figure}[b]
\epsscale{0.8}
\plotone{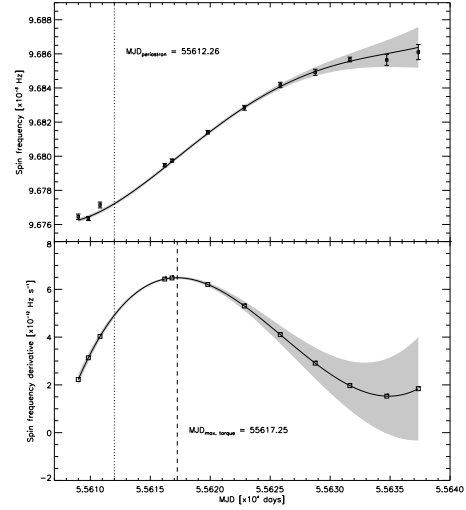}
\caption{
Evolution of the spin frequency ({\it top}) and of the spin frequency derivative ({\it bottom}) of A0535 during the 
outburst. 
Data points in the upper panel show the periods measured with the $Z_2^2$ test.
The solid lines represents the best-fit model obtained from timing analysis. The shaded area denotes the $1\sigma$ error region.
The vertical dotted line shows the epoch of the passage at the periastron.
The dashed vertical line in the lower panel shows the epoch corresponding to the maximum spin frequency derivative 
(i.e. of maximum torque).
}\label{fig_timing}
\end{figure}

We can now infer the evolution of the spin frequency and its first derivative along the outburst (see respectively upper and lower
panels of Figure \ref{fig_timing}).
The increase in the spin frequency induced during the accretion episode is evident.
Moreover, the pulsar was already spinning up during the rising part of the outburst (rev. 1019A to C), 
and it reached a maximum spin-up rate of $(6.48\pm0.05)\times10^{-12}\,\rm Hz\,s^{-1}$ at MJD=55617.25, 
i.e. about five days after the passage at the periastron. 

\begin{figure}[b]
\epsscale{1.0}
\plotone{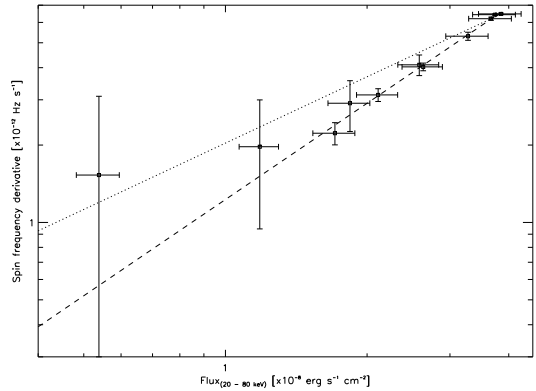}
\caption{Relation between the spin frequency derivative and the 20-80 keV flux of A0535.
Overplotted are two power law models, $\dot{f}\sim(\rm Flux/Flux_{max})^\gamma$, with $\gamma=6/7$ (dotted) and $\gamma=1.24$ 
(dashed), respectivley.\label{fig_torque}}
\end{figure}

The rate of spin-up correlates with the flux of the source (Figure \ref{fig_torque}). 
This behavior is expected in the case of accretion from a prograde accretion disk \citep[e.g.][]{ghosh1979}, because both the 
spin frequency derivative rate and the luminosity are proportional to the accretion rate.
Fitting the $\dot{f}-\rm Flux_{(20-80\,\rm keV)}$ relation with a power law model, $\dot{f}\sim(\rm Flux/Flux_{max})^{\gamma}$, 
we obtain $\gamma = 1.24\pm0.05$, higher than the standard value of 6/7.

\begin{figure*}[b]
\epsscale{1.0}
\plotone{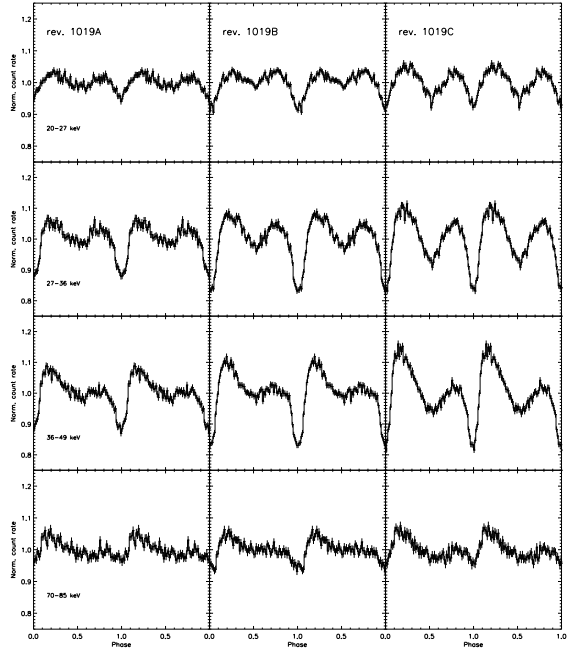}
\caption{Energy-dependent pulse profiles of A0535, normalized to their average value.}.\label{fig_prof_1}
\end{figure*}

\begin{figure*}[b]
\epsscale{1.0}
\plotone{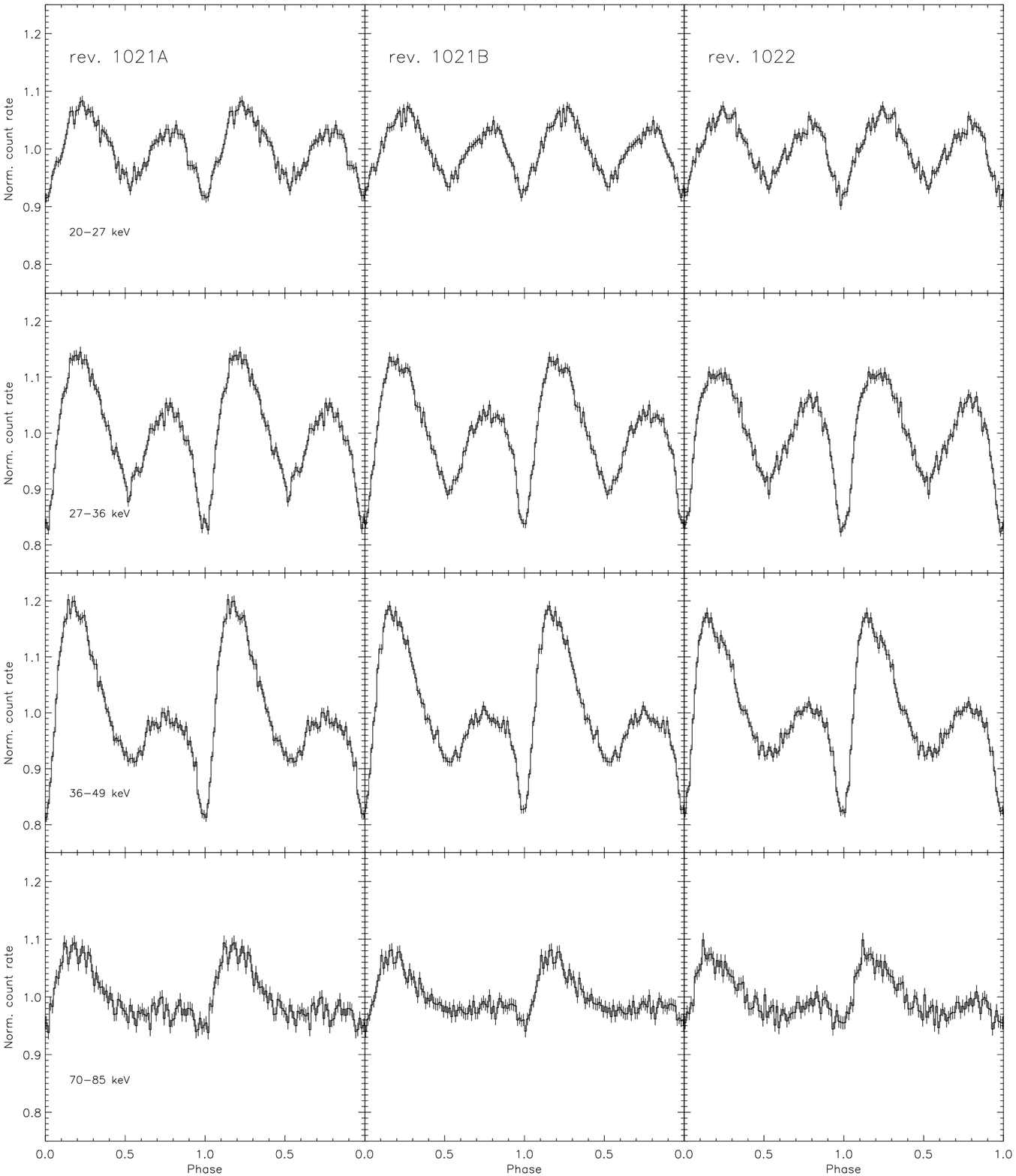}
\caption{...continued from Figure \ref{fig_prof_1}}.\label{fig_prof_2}
\end{figure*}

\begin{figure*}[b]
\epsscale{1.0}
\plotone{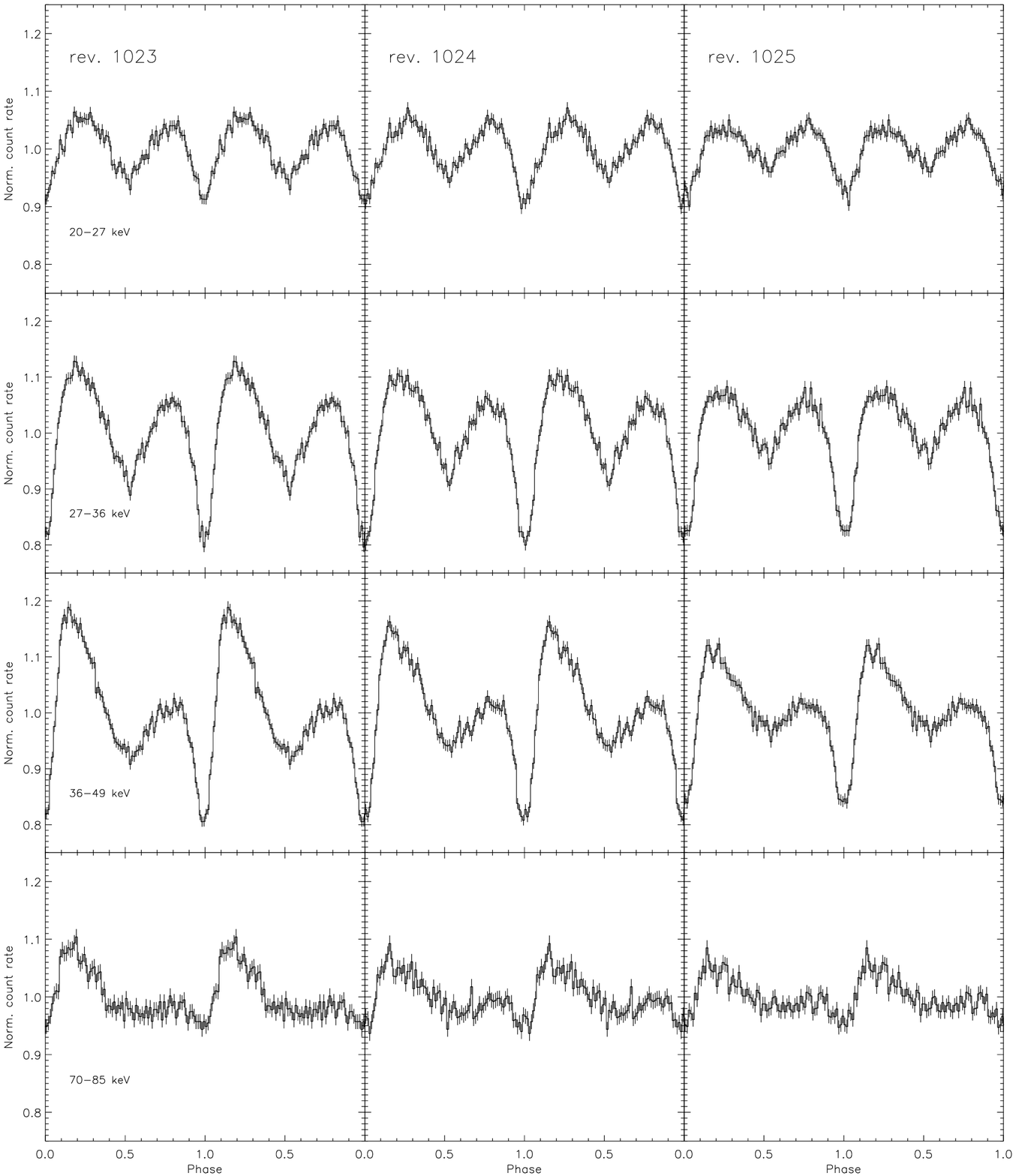}
\caption{...continued from Figure \ref{fig_prof_2}}.\label{fig_prof_3}
\end{figure*}

\begin{figure*}[b]
\epsscale{1.0}
\plotone{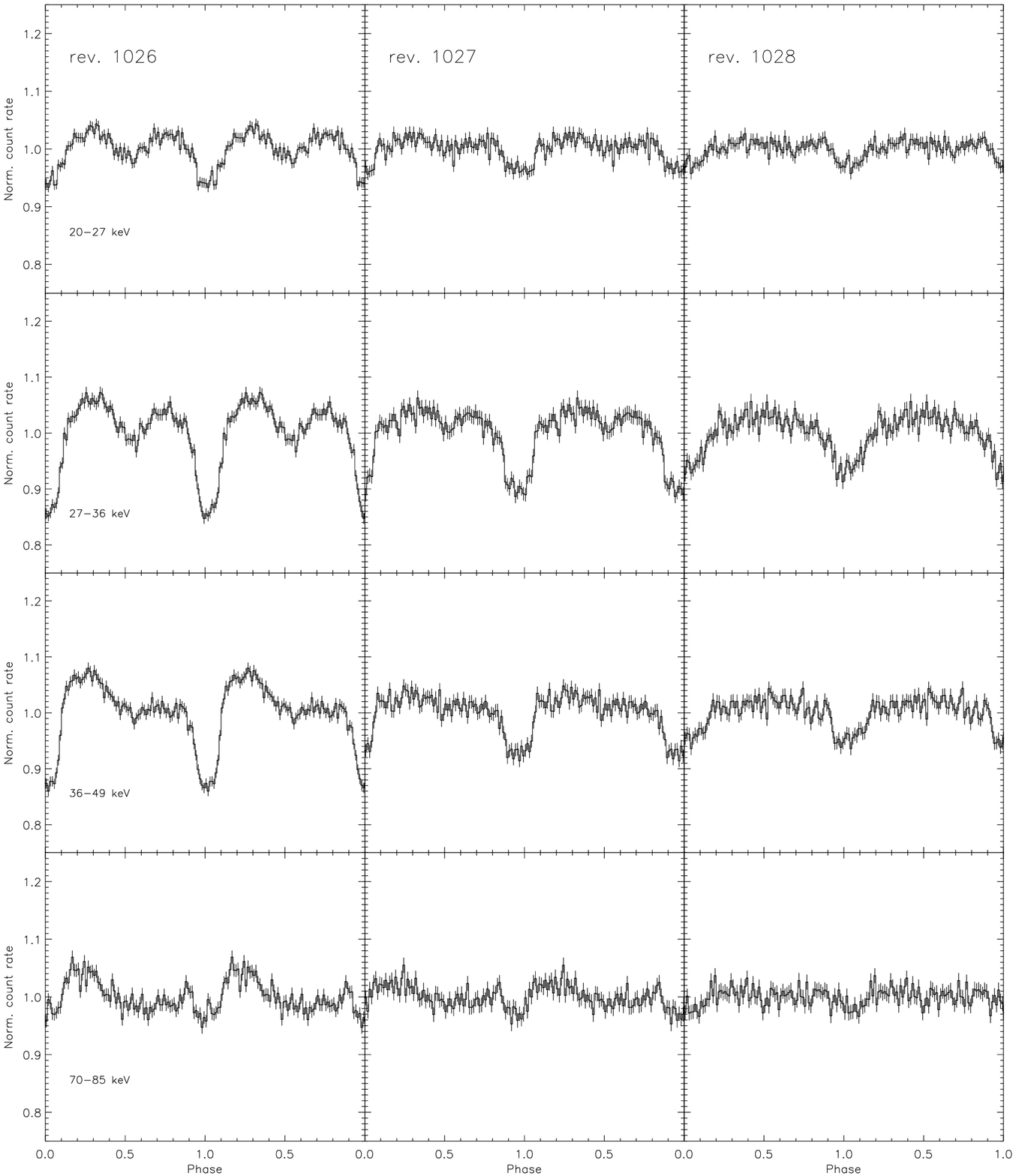}
\caption{...continued from Figure \ref{fig_prof_3}}.\label{fig_prof_4}
\end{figure*}

Finally, we build pulse profiles for each data-set in four different energy bands, $20-27$, $27-36$, $36-49$, and $70-85$ keV, 
respectively, by folding the data according to our estimated timing solution.
The profiles are then normalized to their average value.
We avoid the energy interval between 49 and 70 keV because of the presence of a strong background feature 
\citep[see][]{weidenspointner2003} which significantly reduces the signal-to-noise ratio of the pulsation. 
The resulting pulse profiles are shown in Figures \ref{fig_prof_1} to \ref{fig_prof_4}.
At high luminosity, the pulse profile is made of two sub-pulses. The width of each sub-pulse is $\Delta\phi \sim 0.3$, 
where $\phi$ is the rotational phase in units of the spin period. The relative strength of the sub-pulses is roughly equal 
in the lowest energy range, $20-27\,\rm keV$. 
At higher energies the second peak becomes fainter than the first one, which dominates the pulse profile above 49 keV.
As the luminosity decreases, both sub-pulses weaken. However, it can be noticed that the change in the first sub-pulse is much 
more dramatic than in the second one. 
At energies below 49 keV the profile shows a broad, single pulse ($\Delta\phi \sim 0.8$), while in the $70-85$ keV range 
the lack of statistics does not allow to determine the presence of pulsations.

\section{DISCUSSION}\label{sect-discuss}
We analyzed several {\it INTEGRAL}/SPI observations of the transient Be/X-ray binary A0535, performed during a 
type II outburst of the source occurred in February 2011.

Based on the data taken during a previous outburst, \citet{muller2013} suggested that the lack of significant flux-related
variations of the cyclotron line energy in A0535 result from a peculiar accretion geometry, 
or from a peculiar alignment between the pulsar and an observer at the Earth.
As an alternative, they suggested that A0535 may be in a different accretion regime with respect to other sources, 
where the luminosity is close or below the so-called Coulomb stopping limit \citep[see][for details]{becker2012}, 
and in any case below the critical value $L_{crit}\sim6.8\times10^{37}\,\rm erg\,s^{-1}$, estimated for this source 
at all except the brightest outbursts of the source registered so far.
The results of \cite{caballero2012} for the outburst of February 2011 seem to support this picture.
However, the results of our spectral analysis are in striking contrast with those reported above. 
We found a significant increase in the energy on the fundamental cyclotron line, which correlates with the flux of the source.
This discrepancy might be due to the different energy range used in the spectral analysis, and/or by the different hypotheses
in the modelization of the continuum emission, which might have a considerable effect on the measured energy of the cyclotron lines, 
see e.g. the discussion of \cite{muller2013b}.

In any case, the observation of a positive correlation between the energy of the fundamental cyclotron line and the flux 
confirms that A0535 is accreting at sub-critical regimes. Interestingly, the correlation seems to fall off at the end of the
outburst. This behaviour is expected when the accretion flow passes from a Coulomb-stopping deceleration regime a gas shock 
deceleration regime \citep{becker2012}. In our case, the transition occurs between rev. 1026 and 1027, that is between
$\sim1.2\times10^{37}\,\rm erg\,s^{-1}$ and $\sim6.8\times10^{36}\,\rm erg\,s^{-1}$, in broad agreement with the value
of $\sim7.4\times10^{36}\,\rm erg\,s^{-1}$ obtained from Equation 54 of \cite{becker2012}, where we assumed
for A0535 a magnetic field of $4\times10^{12}\,\rm G$.

The spectral analysis also highlighted a softening of the continuum emission as the flux of the source increased.
A similar behavior of the hard ($20-100\,\rm keV$) X-ray spectrum was also reported by \citet{bildsten1997}
from BATSE observations of the giant outburst of 1994.
This is apparently in contrast with the hardening of the spectrum with flux reported by \citet{muller2013}.
However, the hardness ratio estimated by these authors were obtained from different energy bands, which included the 
soft (5-20 keV) X-ray contribution as well. 
A global hardening of the spectrum could result from a shift of the emission peak towards higher energies. 
This would in turn cause the softening of the spectrum observed above 20 keV.

The spin history of A0535 has been tracked since its discovery \citep[see][and references therein]{camero2012}.
The pulsar shows a steady spin-down during quiescence, which is overlapped with spin-up episodes during giant 
outburst.
From the timing analysis we obtained a phase-connected timing solution for the evolution of the spin frequency relative to the 
February 2011 outburst. Making use of this timing solution, we inferred the evolution of the spin-up rate, which reached a maximum 
value of $\dot{f}=(6.48\pm0.05)\times10^{-12}\, \rm Hz\,s^{-1}$ at MJD 55617.25, in good agreement with the value of
$\dot{f}=(6.35\pm0.05)\times10^{-12}\, \rm Hz\,s^{-1}$ found by \cite{camero2012} using {\it Fermi}/GBM data of the same 
outburst.
We showed that the spin-up rate of A0535 during the outburst was correlated with the flux of the source, 
$\dot{f}\sim(\rm Flux)^{1.24}$, and it is clearly different from the theoretical value of $6/7$ expected in the case of disk 
accretion. 
This discrepancy has been recently reported also by \cite{sugizaki2015} using {\it Fermi}/GBM and MAXI/GSC data of A0535 
and other Be X-ray binaries, and it could be explained by unaccounted variations of the beaming geometry and/or of the bolometric 
correction factor throughout the outburst, which may affect the determination of the true luminosity of the source.

Our study of the energy-dependent pulse profiles and their evolution along the outburst gave results in line with those 
found in previous outbursts of the source \citep{bildsten1997,camero2012}. 
The pulse profiles of A0535 have been analyzed by \cite{caballero2011}, using the beam pattern decomposition method of 
\cite{kraus1995}, and their results are consistent with the emission from a hollow cone (i.e. the accretion column) above the polar 
cap(s) of the neutron star. The transition from a top-hat, energy independent profile at low luminosity to a highly 
asymmetric double-peaked and energy-dependent profile at high luminosity might be explained in terms, for instance, 
of variation of the height and shape of the accretion column. The broad, top-hat and almost energy-independent profile at 
low luminosity suggests that the emission region occupies a large part of the neutron star surface (even accounting for 
light-bending effects) and that the physical conditions, and thus the optical depth, do not change much with the rotational 
phase.
On the other hand, the asymmetric and energy-dependent profile observed at high luminosity might result from a 
non-symmetric accretion column, due to either offset dipole or non-dipolar magnetic fields, and thus, to a non-trivial and 
energy-dependent relation between the physical conditions and geometry of the emission region and the rotational phase.
This relation can be verified by means of phase-resolved spectroscopy, which will be detailed in a forthcoming paper 
(Sartore et al., in preparation).

\acknowledgments
We thank the anonymous referee for helpful comments and suggestions which improved the previous version of the manuscript. 
This work is based on observations of {\it INTEGRAL}, an ESA project with instruments and science data center funded by ESA members 
states (especially the PI countries: Denmark, France, Germany, Italy, Switzerland, Spain), Czech Republic and Poland, 
and with the participation of Russia and the USA. 
NS acknowledges financial support from the French Space Agency CNES through CNRS.

\clearpage

\clearpage

\clearpage

\begin{deluxetable}{c c c}
\tabletypesize{\scriptsize}
\tablecaption{Log of {\it INTEGRAL} observation of A0335+26 during the outburst of February 2011\label{table_log}}
\tablewidth{0pt}
\tablehead{\colhead{Revolution} & \colhead{Start date of the observation} & \colhead{Total exposure}\\
\colhead{} & \colhead{$[\rm YYYY-MM-DD]$} & \colhead{$[\rm ks]$}
}
\startdata
1019 & 2011-02-16 & 186.16  \\
1021 & 2011-02-23 & 101.71  \\
1022 & 2011-02-27 & 53.50  \\
1023 & 2011-03-02 & 43.12  \\
1024 & 2011-03-05 & 43.88  \\
1025 & 2011-03-08 & 46.14  \\
1026 & 2011-03-11 & 64.33  \\
1027 & 2011-03-14 & 50.33  \\
1028 & 2011-03-17 & 50.49  \\
\enddata
\end{deluxetable}

\begin{deluxetable}{c c c c c c c c c c c c c c}
\tabletypesize{\scriptsize}
\rotate
\tablecaption{Best-fit spectral parameters for the {\it INTEGRAL}/SPI observations of A0535.
\label{table_specfit}}
\tablewidth{0pt}
\tablehead{
\colhead{Data-set} & \colhead{1019A} & \colhead{1019B} & \colhead{1019C} & \colhead{1021A} & \colhead{1021B}
& \colhead{1022} & \colhead{1023} & \colhead{1024} & \colhead{1025} & \colhead{1026} & \colhead{1027}
}
\startdata 
$\Gamma$ & 0.5 & 0.5 & 0.5 & 0.5 & 0.5 & 0.5 & 0.5 & 0.5 & 0.5 & 0.5 & 0.5 \\
\\
$E_{cut}$ &
$16.8_{-0.3}^{+0.2}$ & $16.4_{-0.1}^{+0.1}$ & $16.8_{-0.1}^{+0.1}$ &
$15.7_{-0.1}^{+0.1}$ & $15.8_{-0.1}^{+0.2}$ & $15.6_{-0.1}^{+0.1}$ &
$15.9_{-0.1}^{+0.1}$ & $16.5_{-0.1}^{+0.2}$ & $16.4_{-0.2}^{+0.2}$ &
$17.5_{-0.2}^{+0.2}$ & $16.8_{-0.4}^{+0.4}$ \\
\\
$E_{cyc_1}$ & 
$45.6_{-1.0}^{+0.7}$ & $45.6_{-0.7}^{+0.7}$ & $47.4_{-0.8}^{+0.9}$ &
$48.2_{-0.8}^{+0.8}$ & $49.8_{-0.9}^{+1.6}$ & $46.4_{-0.6}^{+0.6}$ &
$46.8_{-0.6}^{+0.7}$ & $47.9_{-0.7}^{+1.1}$ & $46.1_{-1.0}^{+1.0}$ &
$43.9_{-0.4}^{+0.4}$ & $44.9_{-0.9}^{+1.0}$ \\
\\
$\sigma_{cyc_1}$ & 
$12.3_{-1.1}^{+1.6}$ & $8.9_{-0.8}^{+0.9}$ & $10.0_{-1.0}^{+1.1}$ &
$9.8_{-0.8}^{+0.9}$ & $11.1_{-1.0}^{+1.5}$ & $8.8_{-0.6}^{+0.7}$ &
$9.4_{-0.7}^{+0.8}$ & $10.3_{-0.8}^{+1.3}$ & $11.6_{-1.4}^{+1.3}$ &
$9.8_{-0.6}^{+0.7}$ & $9.3_{-1.1}^{+1.3}$ \\
\\
$\tau_{cyc_1}$ & 
$13.1_{-1.9}^{+3.0}$ & $4.9_{-0.6}^{+0.8}$ & $6.3_{-0.9}^{+1.2}$ &
$3.6_{-0.4}^{+0.5}$ & $4.2_{-0.6}^{+1.4}$ & $2.8_{-0.3}^{+0.4}$ &
$3.9_{-0.4}^{+0.6}$ & $5.9_{-0.7}^{+1.6}$ & $8.0_{-1.3}^{+1.9}$ &
$10.3_{-0.9}^{+1.1}$ & $11.6_{-1.9}^{+2.6}$ \\
\\
Significance & 
$>3.9\sigma$ & $>3.9\sigma$ & $>3.9\sigma$ & $>3.9\sigma$ &
$>3.9\sigma$ & $>3.9\sigma$ & $>3.9\sigma$ & $>3.9\sigma$ &
$>3.9\sigma$ & $>3.9\sigma$ & $>3.9\sigma$\\
\\
$E_{cyc_2}$ & 
 - & - & - & - & $98.6_{-1.4}^{+1.7}$ & $107.1_{-1.6}^{+2.0}$ & $105.4_{-2.2}^{+2.2}$ & $100.3_{-3.2}^{+4.6}$ &
 - & $99.2_{-1.6}^{+1.8}$ & - \\
\\
$\sigma_{cyc_2}$ & 
 - & - & - & - & $5.5_{-1.2}^{+1.3}$ & $5.3_{-1.7}^{+1.5}$ & $6.4_{-2.2}^{+1.3}$ & $7.2_{-3.2}^{+3.8}$ &
 - & $4.0_{-2.3}^{+1.0}$\\
\\
$\tau_{cyc_2}$ & 
 - & - & - & - & $17.0_{-2.8}^{+7.4}$ & $23.0_{-4.5}^{+17.7}$ & $29.0_{-6.4}^{+23.7}$ & $10.5_{-2.5}^{+10.2}$ &
 - & $24.7_{-6.3}^{+896.1}$\\
\\
Significance & 
 - & - & - & - & $3.1\sigma$ & $3.6\sigma$ & $>3.9\sigma$ & $2.3\sigma$ & - & $3.5\sigma$ & - \\
\\
Flux & $1.14$ & $1.40$ & $1.76$ & $2.49$ & $2.62$ & $2.46$ & $2.20$ & $1.71$ & $1.21$ & $0.80$ & $0.36$ \\
$[\rm Crab]$ \\
\\
$\chi_{\rm red}^2 (\rm d.o.f.)$ & 
2.34(37) & 1.84(37) & 2.84(37) & 1.20(36) & 1.09(25) & 1.08(34) & 1.41(34) & 1.12(34) & 1.80(37) & 1.50(34) & 1.63(27) \\
\\
\hline
\enddata
\end{deluxetable}

\begin{deluxetable}{c c}
\tabletypesize{\scriptsize}
\tablecaption{Best-fit parameters of the polynomial fit describing the evolution of the spin parameters of A0535 during the 
outburst. Quoted errors are the formal $1\sigma$ uncertainties from the best-fit.\label{table_timing_solution}}
\tablewidth{0pt}
\tablehead{}
\startdata
 Epoch & 55616.202\\
 $[\rm MJD]$ & \\
\\
 $f_0$ & $(9.6793\pm0.0001)\times10^{-3}\, \rm Hz$  \\
\\
 $\dot{f_0}$ & $(6.43\pm0.05)\times10^{-12}\, \rm Hz\,s^{-1}$  \\
\\
 $\ddot{f_0}$ & $(1.21\pm0.17)\times10^{-18}\, \rm Hz\,s^{-2}$  \\
\\
 $\dddot{f_0}$ & $(-1.43\pm0.09)\times10^{-23}\, \rm Hz\,s^{-3}$  \\
\\
 $\ddddot{f_0}$ & $(1.67\pm0.15)\times10^{-29}\, \rm Hz\,s^{-4}$  \\
\enddata
\end{deluxetable}


\begin{thebibliography}{}
\bibitem[Arnaud(1996)]{arnaud1996} Arnaud, K.~A.\ 1996, Astronomical Data Analysis Software and Systems V, 101, 17 
\bibitem[Basko \& Sunyaev(1976)]{basko1976} Basko, M.~M., \& Sunyaev, R.~A.\ 1976, \mnras, 175, 395 
\bibitem[Becker \& Wolff(2007)]{becker2007} Becker, P.~A., \& Wolff, M.~T.\ 2007, \apj, 654, 435 
\bibitem[Becker et al.(2012)]{becker2012} Becker, P.~A., Klochkov, D., Sch{\"o}nherr, G., et al.\ 2012, \aap, 544, A123 
\bibitem[Bildsten et al.(1997)]{bildsten1997} Bildsten, L., Chakrabarty, D., Chiu, J., et al.\ 1997, \apjs, 113, 367 
\bibitem[Buccheri et al.(1983)]{buccheri1983} Buccheri, R., Bennett, K., Bignami, G.~F., et al.\ 1983, \aap, 128, 245 
\bibitem[Burke et al.(2014)]{burke2014} Burke, M.~J., Jourdain, E., Roques, J.-P., \& Evans, D.~A.\ 2014, \apj, 787, 50 
\bibitem[Burnard et al.(1991)]{burnard1991} Burnard, D.~J., Arons, J., \& Klein, R.~I.\ 1991, \apj, 367, 575 
\bibitem[Caballero et al.(2007)]{caballero2007} Caballero, I., Kretschmar, P., Santangelo, A., et al.\ 2007, \aap, 465, L21 
\bibitem[Caballero et al.(2008)]{caballero2008} Caballero, I., Santangelo, A., Kretschmar, P., et al.\ 2008, \aap, 480, L17 
\bibitem[Caballero et al.(2011)]{caballero2011} Caballero, I., Kraus, U., Santangelo, A., Sasaki, M., \& Kretschmar, P.\ 2011, \aap, 526, A131 
\bibitem[Caballero et al.(2012)]{caballero2012} Caballero, I., M{\"u}ller, S., Bordas, P., et al.\ 2012, American Institute of Physics Conference Series, 1427, 300 
\bibitem[Caballero et al.(2013)]{caballero2013} Caballero, I., Pottschmidt, K., Marcu, D.~M., et al.\ 2013, \apjl, 764, L23 
\bibitem[Cabanac et al.(2011)]{cabanac2011} Cabanac, C., Roques, J.-P., \& Jourdain, E.\ 2011, \apj, 739, 58 
\bibitem[Camero-Arranz et al.(2012)]{camero2012} Camero-Arranz, A., Finger, M.~H., Wilson-Hodge, C.~A., et al.\ 2012, \apj, 754, 20 
\bibitem[DeCesar et al.(2013)]{decesar2013} DeCesar, M.~E., Boyd, P.~T., Pottschmidt, K., et al.\ 2013, \apj, 762, 61 
\bibitem[Finger et al.(1996)]{finger1996} Finger, M.~H., Wilson, R.~B., \& Harmon, B.~A.\ 1996, \apj, 459, 288 
\bibitem[Finger et al.(2006)]{finger2006} Finger, M.~H., Camero-Arranz, A., Kretschmar, P., Wilson, C., \& Patel, S.\ 2006, Bulletin of the American Astronomical Society, 38, 359 
\bibitem[F{\"u}rst et al.(2014)]{furst2014} F{\"u}rst, F., Pottschmidt, K., Wilms, J., et al.\ 2014, \apj, 780, 133 
\bibitem[Ghosh \& Lamb(1979)]{ghosh1979} Ghosh, P., \& Lamb, F.~K.\ 1979, \apj, 234, 296 
\bibitem[Jensen et al.(2003)]{jensen2003} Jensen, P.~L., Clausen, K., Cassi, C., et al.\ 2003, \aap, 411, L7 
\bibitem[Jourdain \& Roques(2009)]{jourdain2009} Jourdain, E., \& Roques, J.~P.\ 2009, \apj, 704, 17 
\bibitem[Klochkov et al.(2011)]{klochkov2011} Klochkov, D., Ferrigno, C., Santangelo, A., et al.\ 2011, \aap, 536, L8 
\bibitem[Klochkov et al.(2012)]{klochkov2012} Klochkov, D., Doroshenko, V., Santangelo, A., et al.\ 2012, \aap, 542
\bibitem[Kraus et al.(1995)]{kraus1995} Kraus, U., Nollert, H.-P., Ruder, H., \& Riffert, H.\ 1995, \apj, 450, 763 
\bibitem[Leahy(1987)]{leahy1987} Leahy, D.~A.\ 1987, \aap, 180, 275
\bibitem[Revnivtsev \& Mereghetti(2014)]{revnivtsev2014} Revnivtsev, M., \& Mereghetti, S.\ 2014, \ssr, 58 
\bibitem[Molkov et al.(2010)]{molkov2010} Molkov, S., Jourdain, E., \& Roques, J.~P.\ 2010, \apj, 708, 403 
\bibitem[Mowlavi et al.(2006)]{mowlavi2006} Mowlavi, N., Kreykenbohm, I., Shaw, S.~E., et al.\ 2006, \aap, 451, 187 
\bibitem[M{\"u}ller et al.(2013a)]{muller2013} M{\"u}ller, D., Klochkov, D., Caballero, I., \& Santangelo, A.\ 2013, \aap, 552, A81 
\bibitem[M{\"u}ller et al.(2013b)]{muller2013b} M{\"u}ller, S., Ferrigno, C., K{\"u}hnel, M., et al.\ 2013, \aap, 551, AA6 
\bibitem[Nakajima et al.(2006)]{nakajima2006} Nakajima, M., Mihara, T., Makishima, K., \& Niko, H.\ 2006, \apj, 646, 1125 
\bibitem[Nakajima et al.(2014)]{nakajima2014} Nakajima, M., Mihara, T., Sugizaki, M., et al.\ 2014, \pasj, 66, 9 
\bibitem[Orlandini et al.(2012)]{orlandini2012} Orlandini, M., Frontera, F., Masetti, N., Sguera, V., \& Sidoli, L.\ 2012, \apj, 748, 86 
\bibitem[Porter \& Rivinius(2003)]{porter2003} Porter, J.~M., \& Rivinius, T.\ 2003, \pasp, 115, 1153 
\bibitem[Postnov et al.(2008)]{postnov2008} Postnov, K., Staubert, R., Santangelo, A., et al.\ 2008, \aap, 480, L21 
\bibitem[Postnov et al.(2014)]{postnov2014} Postnov, K.~A., Mironov, A.~I., Lutovinov, A.~A., et al.\ 2014, arXiv:1410.3708 
\bibitem[Ransom et al.(2002)]{ransom2002} Ransom, S.~M., Eikenberry, S.~S., \& Middleditch, J.\ 2002, \aj, 124, 1788 
\bibitem[Roques et al.(2003)]{roques2003} Roques, J.~P., Schanne, S., von Kienlin, A., et al.\ 2003, \aap, 411, L91 
\bibitem[Rosenberg et al.(1975)]{rosenberg1975} Rosenberg, F.~D., Eyles, C.~J., Skinner, G.~K., \& Willmore, A.~P.\ 1975, \nat, 256, 628 
\bibitem[Rothschild et al.(2013)]{rothshild2013} Rothschild, R., Markowitz, A., Hemphill, P., et al.\ 2013, \apj, 770, 19 
\bibitem[Shakura et al.(2012)]{shakura2012} Shakura, N., Postnov, K., Kochetkova, A., \& Hjalmarsdotter, L.\ 2012, \mnras, 420, 216 
\bibitem[Staubert et al.(2007)]{staubert2007} Staubert, R., Shakura, N.~I., Postnov, K., et al.\ 2007, \aap, 465, L25 
\bibitem[Steele et al.(1998)]{steele1998} Steele, I.~A., Negueruela, I., Coe, M.~J., \& Roche, P.\ 1998, \mnras, 297, L5 
\bibitem[Stella et al.(1986)]{stella1986} Stella, L., White, N.~E., \& Rosner, R.\ 1986, \apj, 308, 669 
\bibitem[Sugizaki et al.(2015)]{sugizaki2015} Sugizaki, M., Mihara, T., Nakajima, M., \& Yamaoka, K.\ 2015, arXiv:1502.04461 
\bibitem[Tsygankov et al.(2006)]{tsygankov2006} Tsygankov, S.~S., Lutovinov, A.~A., Churazov, E.~M., \& Sunyaev, R.~A.\ 2006, \mnras, 371, 19 
\bibitem[Vedrenne et al.(2003)]{vedrenne2003} Vedrenne, G., Roques, J.-P., Sch{\"o}nfelder, V., et al.\ 2003, \aap, 411, L63 
\bibitem[Weidenspointner et al.(2003)]{weidenspointner2003} Weidenspointner, G., Kiener, J., Gros, M., et al.\ 2003, \aap, 411, L113 
\bibitem[Wilms(2014)]{wilms2014} Wilms, J.\ 2014, European Physical Journal Web of Conferences, 64, 6001 
\bibitem[Winkler et al.(2003)]{winkler2003} Winkler, C., Courvoisier, T.~J.-L., Di Cocco, G., et al.\ 2003, \aap, 411, L1 
\end{thebibliography}
\end{document}